# Public Sentiment and Demand for Used Cars after A Large-Scale Disaster: Social Media Sentiment Analysis with Facebook Pages

Yuya Shibuya
Graduate School of Interdisciplinary Information Studies,
The University of Tokyo,
Japan,
yuya-shibuya@g.ecc.u-tokyo.ac.jp

Hideyuki Tanaka
Graduate School of Interdisciplinary Information Studies,
The University of Tokyo,
Japan,
tanaka@iii.u-tokyo.ac.jp

## ABSTRACT

*There have been various studies analyzing public sentiment after a large-scale disaster. However, few studies have focused on the relationship between public sentiment on social media and its results on people's activities in the real world. In this paper, we conduct a long-term sentiment analysis after the Great East Japan Earthquake and Tsunami of 2011 using Facebook Pages with the aim of investigating the correlation between public sentiment and people's actual needs in areas damaged by water disasters. In addition, we try to analyze whether different types of disaster-related communication created different kinds of relationships on people's activities in the physical world. Our analysis reveals that sentiment of geo-info-related communication, which might be affected by sentiment inside a damaged area, had a positive correlation with the prices of used cars in the damaged area. On the other hand, the sentiment of disaster-interest-based-communication, which might be affected more by people who were interested in the disaster, but were outside the damaged area, had a negative correlation with the prices of used cars. The result could be interpreted to mean that when people begin to recover, used-car prices rise because they become more positive in their sentiment. This study suggests that, for long-term disaster-recovery analysis, we need to consider the different characteristics of online communication posted by locals directly affected by the disaster and non-locals not directly affected by the disaster.*

## 1 INTRODUCTION

Social data mining in the context of disasters has been widely addressed because of the increasing popularity of social media. However, most studies focus on short-term restoration from disasters and fewer studies analyze social data mining for long-term recovery [25]. In addition, there is a need to study more about the relationships between social media communication and people's actual activities in the real world. With this in mind, this paper analyzes long-term public sentiment after a large-scale disaster and investigates the relationships between public sentiment and people's needs in the disaster-stricken area. More specifically, we compare sentiment in Facebook Page posts and comments related to sentiments about disaster-stricken areas and people as well as used-car price data in the damaged areas of the Great East Japan Earthquake and Tsunami of 2011. The reason we focused on used-car price is that when a water-related large-scale disaster occurs, used cars tend to be increasingly in demand in the damaged areas as shown in subsection 2.1. Our previous study reveals that, after the Great East Japan Earthquake and Tsunami, people in the damaged area needed to buy less expensive cars that required less paperwork for purchase when they began to actively rebuild their daily lives and community, such as when they would go to work, check on their family's safety, and work on recovery efforts [1].

The rest of this paper is constructed as follows. Section 2 provides a literature review and our research questions. In Section 3, we explain the two types of data for our analysis. In Section 4, a sentiment analysis method and its results are described. In Section 5, the model for investigating the relationships between social media sentiment and used-car market data is explained. The result of our analysis is provided in Section 6. Based on this result, we discuss the main findings of our study in Section 7. Lastly, we conclude our study in Section 8.

## 2 RELATED WORK

In this section, we briefly overview related works from two viewpoints: Used-car analysis and social media data analysis focusing on sentiment analysis during disasters.

### 2.1 Used-car demand after the disaster

When a large-scale water-related disaster (e.g., hurricane, typhoon, tsunami) hits a community, used-car demand tends to increase in the community. For example, after Hurricane Harvey and Irma in 2017, the media reported increased demand for used cars around the flooded areas [3, 5, 14]. Likewise, after the Great East Japan Earthquake and Tsunami, newspapers reported a rise in the demand for used cars in the disaster-stricken area to replace water-damaged vehicles and to aid in the rebuilding of the community [20]. [26] statistically shows that there was an increased demand for used cars, particularly for Light Motor Vehicles[1] in the disaster-stricken area compared to less demand in non-damaged areas. Also, [1] suggests that communication on Facebook related to the Great East Japan Earthquake and Tsunami, such as topics of people's activities for recovery and emotional support may have correlated to used-car demand in water-damaged areas. This paper attempts to deepen the understanding of the relationships between online communication and damaged people's needs by applying the Hedonic regression model, which is widely used to assess used-car prices in related studies [1, 10, 22, 26]. The Hedonic Regression Model is a simple economic model that estimates the demand value of a product as reflected in a bundle of embodied characteristics and obtains estimates of the contributory value of each characteristic [22]. By controlling used-car characteristics, the Hedonic model allows us to assess the correlation between used-car prices, which reflect used-car demand [26], and sentiment expressed on Facebook Pages.

## 2.2 Social Media Data Analysis and Sentiment Analysis for Disaster Management

Leveraging social media data before, during and after disasters for disaster management has become popular recently. In particular, social media contents shared by local citizens in a disaster-stricken area can provide unique and critical information for emergency responders, event planners, damaged people, journalists, and digital volunteers [27]. Existing studies have recognized different communication patterns between local and non-locals during disasters [13]. [28] found that twitter users in general were more likely to re-tweet the accounts of people local to emergency events. Moreover, [2] investigated an imaginary representation of a disaster via images shared on Twitter, and found that locals focus more on the business details of the response and the damage in their cities while a non-local population focused more on the images of people suffering.

Another research direction in the context of disasters, which this study also considers, is sentiment analysis. Public mood and emotion after a large-scale disaster have been analyzed broadly [8, 9, 11, 15, 21, 29]. For example, [18] analyzes the tweets during Hurricane Sandy and describes how people's sentiment changed based on their distance from the hurricane. However, analyzing how cyber-world sentiment correlates with physical world data is still in its infancy compared to other fields. In marketing research, correlations between social media sentiment and sales [24], consumer choice [16], and stock prices [23] have been quantitatively analyzed. In public health, the correlations between a pandemic and social media data have been studied with the aim of detecting epidemic outbreaks as soon as possible [4, 6, 7]. There is a need to study more about what cyber-world data during disasters means in a real-world situation. In disaster management research, to the authors' best knowledge, most studies have focused on a relatively short period (e.g., [9, 11, 19]), and no study has revealed the relationships between sentiment expression in the cyber world and people's activities in the real world for long-term recovery. However, the authors assume that monitoring and assessing a community's long-term recovery status by leveraging social media data will aid effective and efficient implementation of recovery efforts because traditional socio-economic recovery indicators (e.g., changes in population, consumption and gross domestic production) are not published in a real-time way. Social media data has the possibility of filling the gap in recovery situational awareness by its timeliness. Therefore, this study investigates public sentiment on social media for recovery phases from a socio-economic perspective, with different groups in mind because local and non-local people tend to express and share information differently.

## 2.3 Research Question

Based on the related works described above we have developed the following research questions;

RQ1: How does public sentiment of disaster-related communication on Facebook relate to the demand for used cars in the damaged area?

RQ2: Do different types of disaster-related communication have different types of relationships to the demand for used cars in the damaged area?

## 3 DATA

To examine the two research questions in 2.3 above, we collected two types of datasets; Facebook Page data and Japanese used-car market data from the damaged area.

## 3.1 Facebook Page dataset

*3.1.1 Facebook Pages Data.* Facebook Pages are places to connect people with the same interests, and they are widely used by businesses, organizations, and individuals. Operators of Facebook Pages can share themed content, information, and activities with users, who can also react, comment, and share these posts with their friends.

Before collecting Facebook Pages' content data, we first selected Facebook Pages located in Japan[2]. As of February 2017, there were 109,046 Facebook Pages in Japan. From these pages, we crawled every post and comment between March 11th and September 11th in 2011 through Facebook Graph API. Through acquiring this data, we found that 16,540 out of 109,046 pages published posts during the period. In total, 873,005 posts and comments were gathered. Some of these posts and comments also have attachment files, such as notes, photos, videos, and links to other websites. In this study, we analyzed every text data including attachment descriptions.

*3.1.2 Disaster-related posts.* To examine whether different types of disaster-related communication have different types of effects on the demand for used cars, we created the following four datasets.

*Dataset1) Geo-information based dataset.* First, we selected posts, comments, and descriptions which contained any of the tsunami-stricken areas' names, including cities, towns, villages, and small district's names (totally 2,092 areas' names)[3]. In this method, we found 67,330 posts and comments.

---

[1] A Japanese category of light motor vehicles whose engine volumes are 660cc or less.

[2] The initial list of the Facebook Pages was provided by Professor Shyhtsun Felix Wu, the University of California, Davis.

*Dataset2) Disaster-interests-based dataset.* Second, we tested word appearance proportions between 2010 (before the disaster) and 2011 (after the disaster). In doing so, we further collected Facebook Pages in Japan between March 11th and September 11th in 2010 and found 220,871 posts and comments. By conducting a chi-square test of nouns, adjectives, and verbs whose letter lengths are more than one (in Japanese), we identified 3,385 words whose appearance ratios were significantly larger than those in 2010 ($p < .05$). Because we found several non-disaster-related words among the 3,385 words, for the purpose of this study, the authors manually picked 623 words from the 3,385 words if a word related to the Great East Japan Earthquake and Tsunami and automobile demand. In total, 264,441 posts and comments were found.

*Dataset3)* In addition, we created an intersection of Dataset1 and Dataset2 as Dataset3 (56,017 posts/comments).

*Dataset4)* contains data which consists of only interest-based disaster-related posts and comments (subtracting Dataset3 from Dataset2). Figure 1 shows the number of posts and comments in each dataset and the relationships between them.

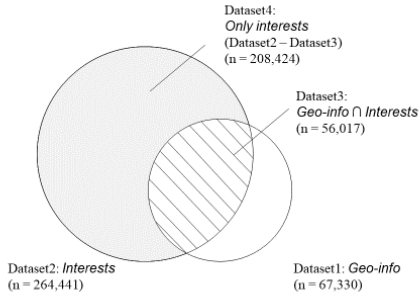

**Figure 1: Four types of disaster-related Facebook Page datasets and number of posts and comments of each dataset**

## 3.2 Used-car market data

The second data we used for the analysis was Japanese used-car market data. We used used-car data sourced from advertisements which were posted on one of the most dominant used-car magazines in Japan, 'Goo.' 'Goo' is published half-monthly. We used the data posted by dealers in the water-damaged areas (Miyagi and Iwate prefectures), between the second half of April in 2011 and the first half of October in 2011. After the disaster, 'Goo' suspended publication of the issue for the first half of April in the damaged area. In total, we analyzed data from 12 issues covering half of the year after the disaster. For the analysis, we selected four types of Light Motor Vehicle used cars because the demand for Light Motor Vehicles notably increased in the damaged area after the disaster [26]. Table 1 describes real prices and the numbers of these body types[4].

**Table 1: Real prices and numbers in the damaged area between the second half of April and the first half of October in 2011**

| Body type | Number of cars | Real Price (yen) | | | |
|---|---|---|---|---|---|
| | | Min | Max | Mean | S.D. |
| Light Motor Vehicle RV (LR) | 16,333 | 49,281 | 2,975,359 | 626,484 | 314,974 |
| Light Motor Vehicle Cab Van (LC) | 1,314 | 59,609 | 1,469,681 | 538,981 | 258,629 |
| Light Motor Vehicle Truck (LT) | 1,280 | 71,942 | 1,565,262 | 482,745 | 246,275 |
| Light Motor Vehicle Others (LO) | 3,567 | 39,014 | 1,745,118 | 520,550 | 280,973 |

## 4. Facebook Page Sentiment Analysis

To analyze the sentiment of each Facebook Page's posts and comments, we have also used Japanese sentiment porality dictionaries designed to use for sentiment analysis, which are openly provided by Inui-Okazaki Laboratory, Tohoku University [12, 17]. The dictionaries contain about 5,000 verbs, adjectives, and adjectival verbs, and about 8,500 nouns. Every word in the dictionaries is tagged as positive, negative, or neutral. For the sake of this study, we removed ten disaster-related words from the dictionaries; namely, 'disaster-stricken,' 'disaster,' 'Tsunami,' 'evacuation,' 'earthquake,' 'catastrophe,' 'flooded,' 'disaster-stricken area,' and 'big disaster.' With the dictionaries, we enumerated the numbers of positive, negative, and neutral words in each post and comment. Then, the sentiment score of the $i$ th post or comment is calculated as $S_i$;

$$\frac{N_i^{Positive} - N_i^{Negative}}{N_i^{Positive} + N_i^{Negative} + N_i^{Neutral}}$$

where $N_i^{Positive}$, $N_i^{Negative}$ and $N_i^{Neutral}$ denote the number of positive, negative and neutral words of the $i$ th comment or post[5]. If the $i$ th post or comment had any attachment description and the attachment description was different from the $i$ th text, the attachment description's score was added to $S_i$.

To analyze these scores with used-car prices, the means of sentiment score in every half month was calculated. Figure 2 shows the chronological change of the sentiment scores means for Dataset1, Dataset2, Dataset3, and Dataset4, respectively. The trends of Dataset 1 and 3 and those of Dataset 2 and 4 are different and sometimes go in opposite directions as the figure shows.

---

[3] This study used areas' name based on 2015 Population Census (https://www.e-stat.go.jp/). If a city, town, or village is listed in the 'Area of inundated area' published by the Geospatial Information Authority of Japan (http://www.gsi.go.jp/kikaku/kikaku60004.html) and located in Miyagi and Iwate prefectures, we regard the area as tsunami-stricken.

[4] Real prices of the used cars were calculated based on the monthly 'automobile' deflator of the fiscal 2015 Consumer Price index (CPI).
[5] When polar words appear with denial words in one phrase (e.g., I was not happy, where 'happy' is a positive word and 'not' is a denial.), the word (e.g., happy) is counted as an opposite polar word (e.g., happy as negative).

For the analysis, we set a one-month time lag between Facebook Pages' data and used-car data because Facebook data is real-time information, but used-car data is paper-based and takes about one month to be published after data was collected from town used-car shops. For example, Facebook Page data of the second half of April 2011 corresponds to the used-car data of the second half of May 2011 in this study. Our target time range was from the second half of April to the second half of September in 2011.

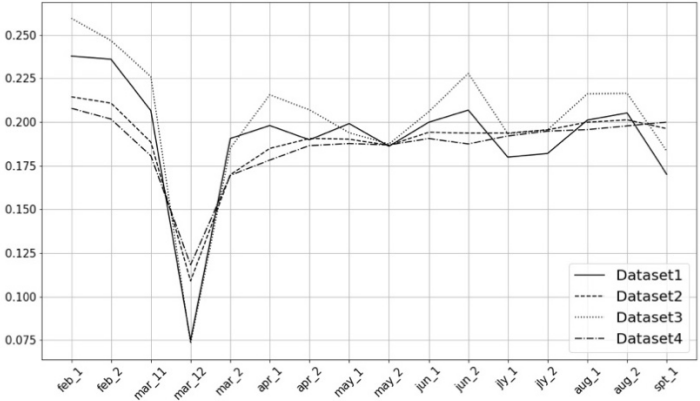

Note: 1 and 2 in the x-axis label represent the first half or second half of the month. Because the disaster happened on the 11th of March 2011, the first half of March in 2011 is divided into two: the first half before the disaster (March 1st to March 10th), and the first half after the disaster (March 11th to 31st).

**Figure 2: Sentiment Score mean of Dataset1-4**

## 5. MODEL

To assess how $S$ correlates with used-car prices, we developed our model, based on the Hedonic Model used in our previous work [10]:

$$\ln P_j = \beta_0 + \beta_1 X_j + \beta_2 S_j + \beta_3 D_j + \varepsilon_j \quad (1)$$

Where $\ln P_j$ is the natural logarithm of the real price of the $j$ th product, $X_j$ is a vector of observable characteristics of the used cars. $S_j$ is Facebook Pages' sentiment scores. $D_j$ is a dummy vector of periods. We chose three periods in order to control chronological effect on prices[6]. $\varepsilon_j$ is the error term. By focusing on $S_j$, we can assess the correlation between the sentiment scores on disaster-related posts and comments on Facebook Pages and the price of used cars in the damaged area by controlling the observable characteristics of the used cars and the chronological effect based on the hedonic approach. For the control variables $X_j$,

this paper uses the following[7]:

Transmission:
$X_{1j}$ = Transmission dummy (Automatic = 1, others = 0)
Fuel:
$X_{2j}$ = Diesel dummy (Diesel = 1, others =0)
$X_{3j}$ = Gas Hybrid dummy (Gas Hybrid = 1, others = 0)
$X_{4j}$ = EV dummy (EV = 1, others = 0)
$X_{5j}$ = Other fuels dummy (LPG, CNG or FC = 1, others = 0)
Age:
$X_{6j}$ = Age (in years)
Kilometers driven:
$X_{7j}$ = 100,000km dummy (over 100k km driven = 1, others = 0)

We apply equation (1) to four types of datasets respectively. The statistical summary and correlation tables of our model are shown in the Appendix.

**Table 2. The result of our models**

| | | | Adjr² | Intercept | X1 | X2 | X3 | X4 | X5 | X6 | X7 | D_72 | D_82 | D_91 | S |
|---|---|---|---|---|---|---|---|---|---|---|---|---|---|---|---|
| Dataset1 | LO | coef | .666 | 0.0 | -0.043** | 0.0 | 0.0* | 0.0* | 0.0* | -0.129** | -0.296** | -0.019 | -0.002 | -0.048* | **1.651*** |
| | | std err | | 0.0 | 0.014 | 0.0 | 0.0 | 0.0 | 0.0 | 0.002 | 0.019 | 0.024 | 0.022 | 0.023 | 0.663 |
| | LC | coef | .700 | 0.0** | 0.048** | 0.0 | 0.0 | 0.0 | 0.102 | -0.082** | -0.357** | -0.033 | -0.020 | -0.048 | 0.801 |
| | | std err | | 0.0 | 0.018 | 0.0 | 0.0 | 0.0 | 0.144 | 0.002 | 0.018 | 0.032 | 0.031 | 0.028 | 0.857 |
| | LR | coef | .660 | 0.0 | 0.092** | 0.318* | 0.267 | -0.071 | 0.309** | -0.10** | -0.321** | -0.006 | 0.013 | -0.015 | **0.942**** |
| | | std err | | 0.0 | 0.005 | 0.157 | 0.182 | 0.182 | 0.087 | 0.001 | 0.007 | 0.010 | 0.009 | 0.009 | 0.255 |
| | LT | coef | .636 | 0.0** | 0.229** | 0.0 | 0.0 | 0.0 | 0.0 | -0.078** | -0.343** | -0.007 | -0.017 | -0.016 | -0.095 |
| | | std err | | 0.0 | 0.038 | 0.0 | 0.0 | 0.0 | 0.0 | 0.002 | 0.027 | 0.035 | 0.035 | 0.033 | 1.009 |
| Dataset2 | LO | coef | .665 | 0.0** | -0.042** | 0.0 | 0.0 | 0.0 | 0.0 | -0.129** | -0.295** | 0.011 | -0.010 | -0.016 | -1.964 |
| | | std err | | 0.0 | 0.014 | 0.0 | 0.0 | 0.0 | 0.0 | 0.002 | 0.019 | 0.021 | 0.022 | 0.024 | 1.347 |
| | LC | coef | .703 | 0.0** | 0.043* | 0.0** | 0.0** | 0.0** | 0.139 | -0.082** | -0.359** | -0.008 | -0.003 | 0.011 | **-6.240**** |
| | | std err | | 0.0 | 0.018 | 0.0 | 0.0 | 0.0 | 0.144 | 0.002 | 0.018 | 0.029 | 0.030 | 0.030 | 1.689 |
| | LR | coef | .660 | 0.0** | 0.092** | 0.325* | 0.266 | -0.078 | 0.310** | -0.10** | -0.321** | 0.011 | 0.009 | 0.005 | **-1.465**** |
| | | std err | | 0.0 | 0.005 | 0.158 | 0.182 | 0.182 | 0.087 | 0.001 | 0.007 | 0.009 | 0.009 | 0.009 | 0.557 |
| | LT | coef | .636 | 0.0** | 0.228** | 0.0 | 0.0 | 0.0 | 0.0 | -0.078** | -0.343** | -0.012 | -0.023 | -0.032 | 1.761 |
| | | std err | | 0.0 | 0.038 | 0.0 | 0.0 | 0.0 | 0.0 | 0.002 | 0.027 | 0.031 | 0.035 | 0.033 | 1.641 |
| Dataset3 | LO | coef | .666 | 0.0** | -0.042** | 0.0* | 0.0* | 0.0** | 0.0** | -0.129** | -0.296** | -0.037 | -0.010 | -0.057* | **1.595*** |
| | | std err | | 0.0 | 0.014 | 0.0 | 0.0 | 0.0 | 0.0 | 0.002 | 0.019 | 0.027 | 0.022 | 0.024 | 0.621 |
| | LC | coef | .700 | 0.0** | 0.048** | 0.0 | 0.0 | 0.0 | 0.104 | -0.082** | -0.357** | -0.022 | -0.027 | -0.041 | 0.079 |
| | | std err | | 0.0 | 0.018 | 0.0 | 0.0 | 0.0 | 0.145 | 0.002 | 0.018 | 0.036 | 0.030 | 0.030 | 0.787 |
| | LR | coef | .660 | 0.0** | 0.092** | 0.320** | 0.268 | -0.073 | 0.310** | -0.10** | -0.321** | 0.011 | 0.009 | 0.008 | **.684**** |
| | | std err | | 0.0 | 0.005 | 0.158 | 0.182 | 0.182 | 0.087 | 0.001 | 0.007 | 0.011 | 0.008 | 0.009 | 0.245 |
| | LT | coef | .636 | 0.0** | 0.230** | 0.0 | 0.0 | 0.0 | 0.0 | -0.078** | -0.343** | -0.016 | -0.015 | -0.022 | 0.265 |
| | | std err | | 0.0 | 0.038 | 0.0 | 0.0 | 0.0 | 0.002 | 0.027 | 0.040 | 0.034 | 0.035 | 0.926 | |
| Dataset4 | LO | coef | .666 | 0.0** | -0.043** | 0.0* | 0.0** | 0.0** | 0.0** | -0.129** | -0.294** | 0.0 | -0.003 | -0.014 | **-2.896**** |
| | | std err | | 0.0 | 0.014 | 0.0 | 0.0 | 0.0 | 0.0 | 0.002 | 0.019 | 0.021 | 0.022 | 0.023 | 1.093 |
| | LC | coef | .703 | 0.0** | 0.043* | 0.0** | 0.0 | 0.0** | 0.136 | -0.082** | -0.359** | -0.030 | 0.0 | -0.009 | **-5.007**** |
| | | std err | | 0.0 | 0.018 | 0.0 | 0.0 | 0.0 | 0.144 | 0.002 | 0.018 | 0.029 | 0.030 | 0.028 | 1.369 |
| | LR | coef | .660 | 0.0** | 0.092** | 0.324** | 0.261* | -0.082 | 0.310** | -0.10** | -0.321** | 0.004 | 0.013 | -0.006 | **-1.743**** |
| | | std err | | 0.0 | 0.005 | 0.157 | 0.182 | 0.182 | 0.087 | 0.001 | 0.007 | 0.009 | 0.009 | 0.009 | 0.447 |
| | LT | coef | .660 | 0.0** | 0.228** | 0.0 | 0.0 | 0.0 | 0.0 | -0.078** | -0.343** | -0.006 | -0.023 | -0.026 | 1.316 |
| | | std err | | 0.0 | 0.038 | 0.0 | 0.0 | 0.0 | 0.002 | 0.027 | 0.031 | 0.035 | 0.033 | 1.641 | |

Note1: Note: D_72, D_82, D_91 represent dummies of the first half of July, the second half of August, and the first half of September, respectively. This applies to the tables below.
Note2: $P$ value *: < .05, **: < .01

## 6. RESULT

By applying equation (1) to four types of datasets, this study observes how public sentiment toward the disaster correlates with used-car price, and whether different types of communication about the disaster have a different correlation with the used-car prices. Table 2 shows the result of our analysis. The estimated coefficients of $S$ are bolded if they have a statistically significant effect ($p < .05$).

For Dataset1 (*Geo-info*), there were significant positive correlations of $S$ with the price of LR and LO. Similarly, in Dataset3 (*Geo-info and disaster-related interests*), there was a significant positive effect of $S$ on the price of LR and LO. On the other hand, in Dataset2 (*Only Disaster-related interests*), there were significant negative correlations of $S$ for the price of LR and LC. In addition, for Dataset4 (*Disaster-related interests*), there were significant negative correlations of $S$ for LR, LC, and LO.

---

[6] The first half of July, the second half of August, and the first half of September are chosen as control periods. We assessed all the periods as a dummy variables based on equation (1) by using all-type-pooling data. Those of the three periods are statistically significantly positive.

[7] Related research uses engine volume as another control variable, but the current study does not because all target light motor vehicles' engine volumes are the same (660cc).

## 7. DISCUSSION

In this section, we discuss the results presented in the previous section. First our analysis reveals that there was a statistically significant positive correlation between Facebook sentiment scores and the price of LR and LC in geo-info based datasets – Dataset1 and Dataset3. Therefore, when geo-info-based disaster-related communication was more positive, there might have been excess demand for LR and LO in the damaged area. On the other hand, our analysis also shows that there were significant negative correlations between Facebook sentiment scores and the prices of several used-car types in disaster-related interests-based datasets – Dataset2 and Dataset4. These negative correlations between sentiment scores and used-car prices might be affected by the different trends of sentiment scores between geo-info-based and disaster-interested-based communication. There are two interpretations of this result. One interpretation is that when people's sentiments in the damaged area became positive, they started to purchase used cars. The other interpretation is that when people are just beginning to recover as indicated in rising car prices, they become more positive in their sentiment. The latter supports the idea that people do not remain helpless victims of disaster response, and instead participate in recovery in ways that are helpful.

For further discussion, we focus on a comparison of Dataset3 and Dataset4 because both are equally related to disaster-interests (Figure 1). Facebook Pages users whose post or comment categorized in Dataset3 might be familiar with damaged areas because they know the names of damaged-areas while those of Dataset4 might not be familiar with these areas. Moreover, the trends of sentiment scores are different. In particular, those of Dataset3 and Dataset4 after the first half of April are negatively correlated (coefficient = -0.28). This difference might imply that the sentiment of those who are familiar with damaged areas such as local residents might change in the opposite direction to the sentiment of those who are not familiar with them such as non-local residents. This means, for instance, that even when people's activities in the damaged area aimed at restoring their daily-lives were intense and their sentiment positively improved, people outside the damaged area might express their anxiety or concern due to reading comments about people's continuous struggles for recovery in the damaged area. These results are consistent with other research that indicates that local citizens more likely share images related to response, recovery and the damage, while the non-local population focused more on the images of people suffering [2]. In long-term recovery, local residents might have had more positive sentiments regarding disaster than those on the outside. This is because they need to be strong to return back to normal routines while onlookers from the outside might continue to express sentiments of worry as part of their collective gaze on others' suffering [2]. Our results imply that, for long-term recovery analysis, we need to distinguish the sentiment differences between those who familiar with the damaged areas and those who are not familiar with the areas even if both are interested in the disaster recovery.

## 8. CONCLUSION

The purpose of this study was to analyze the following two questions; (1) how the public sentiment of disaster-related communication on Facebook related to the demand for used cars in the damaged area, and (2) whether different types of disaster-related communication related differently to the demand for used cars.

For the first question, our analysis reveals that there were correlations between social media sentiment and the price of used cars in the damaged area. For the latter question, this study shows that a correlation between disaster-related communication and used-car demand might depend on whether the people communicating are local or non-local. When those who are familiar with the disaster areas, such as local residents, communicate about the disaster, their sentiment on Facebook Pages might have positively correlated with the price of used cars. One interpretation of this result is that when people are beginning to recover as indicated in rising car prices, they become more positive in their sentiment.

On the other hand, the sentiment of those who are not familiar with the disaster areas, i.e. non-local people might negatively correlate to the used-car demand. This results suggest that we need to distinguish sentiment between those who are familiar with the damaged area and those who are not familiar even if both are equally interested in the disaster itself and disaster relief.

Our study was able to contribute academically to disaster management with the following two points. First, this study showed the potential usefulness of social media data for assessing long-term disaster recovery status in disaster-stricken areas. Secondly, we were able to shed light on a new point of discussion; namely, whether for long-term recovery different types of disaster-related communication on social media have different types of effects in the real world.

On the other hand, we only investigate one disaster. There is a need to investigate other water-related large-scale disasters, such as hurricanes. In addition, future studies should investigate whether sentiment, in general, helps determine other types of public demands other than those of used cars.

## APPENDIX

**Table A1 Statistical summary of the models**

| | | LnP | X1 | X2 | X3 | X4 | X5 | X6 | X7 | D 72 | D 82 | D 91 | Data set1 S | Data set2 S | Data set3 S | Data set4 S |
|---|---|---|---|---|---|---|---|---|---|---|---|---|---|---|---|---|
| LA N=3,567 | mean | 12.99 | 0.72 | 0.00 | 0.00 | 0.00 | 0.00 | 7.61 | 0.14 | 0.10 | 0.10 | 0.09 | 0.19 | 0.19 | 0.20 | 0.19 |
| | std | 0.65 | 0.45 | 0.00 | 0.00 | 0.00 | 0.00 | 3.81 | 0.34 | 0.30 | 0.29 | 0.29 | 0.01 | 0.01 | 0.01 | 0.01 |
| | min | 10.57 | 0.00 | 0.00 | 0.00 | 0.00 | 0.00 | 0.00 | 0.00 | 0.00 | 0.00 | 0.00 | 0.17 | 0.17 | 0.18 | 0.17 |
| | max | 14.37 | 1.00 | 0.00 | 0.00 | 0.00 | 0.00 | 20.00 | 1.00 | 1.00 | 1.00 | 1.00 | 0.21 | 0.20 | 0.23 | 0.20 |
| LC N=1,314 | mean | 13.07 | 0.29 | 0.00 | 0.00 | 0.00 | 0.00 | 7.37 | 0.31 | 0.09 | 0.08 | 0.10 | 0.19 | 0.19 | 0.20 | 0.19 |
| | std | 0.53 | 0.45 | 0.00 | 0.00 | 0.00 | 0.06 | 4.25 | 0.46 | 0.28 | 0.27 | 0.30 | 0.01 | 0.01 | 0.01 | 0.01 |
| | min | 11.00 | 0.00 | 0.00 | 0.00 | 0.00 | 0.00 | 0.00 | 0.00 | 0.00 | 0.00 | 0.00 | 0.17 | 0.17 | 0.18 | 0.17 |
| | max | 14.20 | 1.00 | 0.00 | 0.00 | 0.00 | 1.00 | 21.00 | 1.00 | 1.00 | 1.00 | 1.00 | 0.21 | 0.20 | 0.23 | 0.20 |
| LR N=16,333 | mean | 13.21 | 0.58 | 0.00 | 0.00 | 0.00 | 0.00 | 7.40 | 0.16 | 0.09 | 0.10 | 0.10 | 0.19 | 0.19 | 0.20 | 0.19 |
| | std | 0.54 | 0.49 | 0.02 | 0.01 | 0.01 | 0.03 | 3.73 | 0.37 | 0.29 | 0.30 | 0.30 | 0.01 | 0.00 | 0.01 | 0.01 |
| | min | 10.81 | 0.00 | 0.00 | 0.00 | 0.00 | 0.00 | 0.00 | 0.00 | 0.00 | 0.00 | 0.00 | 0.17 | 0.17 | 0.18 | 0.17 |
| | max | 14.91 | 1.00 | 1.00 | 1.00 | 1.00 | 1.00 | 21.00 | 1.00 | 1.00 | 1.00 | 1.00 | 0.21 | 0.20 | 0.23 | 0.20 |
| LT N=1,280 | mean | 12.95 | 0.06 | 0.00 | 0.00 | 0.00 | 0.00 | 10.00 | 0.14 | 0.10 | 0.08 | 0.10 | 0.19 | 0.19 | 0.20 | 0.19 |
| | std | 0.54 | 0.24 | 0.00 | 0.00 | 0.00 | 0.00 | 5.27 | 0.35 | 0.30 | 0.27 | 0.30 | 0.01 | 0.01 | 0.01 | 0.01 |
| | min | 11.18 | 0.00 | 0.00 | 0.00 | 0.00 | 0.00 | 0.00 | 0.00 | 0.00 | 0.00 | 0.00 | 0.17 | 0.17 | 0.18 | 0.17 |
| | max | 14.26 | 1.00 | 0.00 | 0.00 | 0.00 | 0.00 | 22.00 | 1.00 | 1.00 | 1.00 | 1.00 | 0.21 | 0.20 | 0.23 | 0.20 |

## Table A2 Correlation table of the model for Each Dataset

| | | Dataset2 | | | | | | | | | | |
|---|---|---|---|---|---|---|---|---|---|---|---|---|
| | | LnP | X1 | X2 | X3 | X4 | X5 | X6 | X7 | D_72 | D_82 | D_91 | S |
| Dataset1 | LnP | 1 | .137 | .216 | .127 | -.007 | .003 | -.497 | -.261 | .0 | .016 | .009 | .022 |
| | X1 | .137 | 1 | -.080 | .053 | -.002 | -.017 | -.166 | -.067 | -.002 | .008 | .002 | .031 |
| | X2 | .216 | -.080 | 1 | -.024 | -.002 | -.005 | .217 | .210 | -.003 | -.003 | -.003 | -.022 |
| | X3 | .127 | .053 | -.024 | 1 | -.001 | -.002 | -.085 | -.019 | -.004 | .005 | .004 | .012 |
| | X4 | -.007 | -.002 | -.002 | -.001 | 1 | .0 | .004 | -.005 | -.003 | .002 | -.003 | -.009 |
| | X5 | .003 | -.017 | -.005 | -.002 | .0 | 1 | .002 | .004 | -.004 | -.004 | -.001 | .0 |
| | X6 | -.497 | -.166 | .217 | -.085 | .004 | .002 | 1 | .297 | -.003 | -.020 | -.012 | -.053 |
| | X7 | -.261 | -.067 | .210 | -.019 | -.005 | .004 | .297 | 1 | -.006 | -.006 | -.002 | -.009 |
| | D_72 | .0 | -.002 | -.003 | -.004 | -.003 | -.004 | -.003 | -.006 | 1 | -.102 | -.103 | .013 |
| | D_82 | .016 | .008 | -.003 | .005 | .002 | -.004 | -.020 | -.006 | -.102 | 1 | -.106 | .130 |
| | D_91 | .009 | .002 | -.003 | .004 | -.003 | -.001 | -.012 | -.002 | -.103 | -.106 | 1 | .413 |
| | S | -.008 | -.010 | -.003 | -.011 | -.002 | .003 | .011 | -.004 | .40 | -.289 | .252 | - |

| | | Dataset4 | | | | | | | | | | |
|---|---|---|---|---|---|---|---|---|---|---|---|---|
| | | LnP | X1 | X2 | X3 | X4 | X5 | X6 | X7 | D_72 | D_82 | D_91 | S |
| Dataset3 | LnP | 1 | .137 | .216 | .127 | -.007 | .003 | -.497 | -.261 | .0 | .016 | .009 | .025 |
| | X1 | .137 | 1 | -.080 | .053 | -.002 | -.017 | -.166 | -.067 | -.002 | .008 | .002 | .035 |
| | X2 | .216 | -.080 | 1 | -.024 | -.002 | -.005 | .217 | .210 | -.003 | -.003 | -.003 | -.017 |
| | X3 | .127 | .053 | -.024 | 1 | -.001 | -.002 | -.085 | -.019 | -.004 | .005 | .004 | .016 |
| | X4 | -.007 | -.002 | -.002 | -.001 | 1 | .0 | .004 | -.005 | -.003 | .002 | -.003 | -.007 |
| | X5 | .003 | -.017 | -.005 | -.002 | .0 | 1 | .002 | .004 | -.004 | -.004 | -.001 | .0 |
| | X6 | -.497 | -.166 | .217 | -.085 | .004 | .002 | 1 | .297 | -.003 | -.020 | -.012 | -.054 |
| | X7 | -.261 | -.067 | .210 | -.019 | -.005 | .004 | .297 | 1 | -.006 | -.006 | -.002 | -.006 |
| | D_72 | .0 | -.002 | -.003 | -.004 | -.003 | -.004 | -.003 | -.006 | 1 | -.102 | -.103 | -.181 |
| | D_82 | .016 | .008 | -.003 | .005 | .002 | -.020 | -.006 | -.102 | 1 | -.106 | .205 | |
| | D_91 | .009 | .002 | -.003 | .004 | -.003 | -.001 | -.012 | -.002 | -.103 | -.106 | 1 | .253 |
| | S | -.002 | -.008 | -.007 | -.007 | -.003 | .0 | .0 | -.005 | .561 | -.190 | .30 | - |

## ACKNOWLEDGEMENTS

We thank the reviewers of this paper for their constructive feedback including insightful suggestion of an interpretation of our results. Any errors are the responsibility of the authors. This study was partially supported by the research grant 'Proto Award,' which the first author received. We would like to show our gratitude to Proto Corporation who provided used-car market data and financial support. This study is also partially supported by the Graduate Program for Social ICT Global Creative Leaders. The authors also would like to thank Professor Shyhtsun Felix Wu from the University of California, Davis, for providing us with the initial data of the Japanese Facebook Pages' profiles. This study received ethical approval (No.17-2) from the Graduate School of Interdisciplinary Information Studies, The University of Tokyo.